\documentclass{mn2e}
\usepackage{times}

\input{psfig.sty}

\newif\ifAMStwofonts

\AMStwofontstrue

\def\37{NGC~3783}
\def\55{NGC~5548}
\def\74{NGC~7469}

\def\p{$\pm$}
\def\etal{{\it et~al.~}}

\title{The complex iron line in NGC~7469 observed by BeppoSAX}
\author[A. De Rosa et al.]
       {A. De Rosa, $^{1,2}$ A. C. Fabian $^1$ and L. Piro $^2$ \\
        $^1$Institute of Astronomy, 
	    Madingley Road, Cambridge CB3 0HA \\
	$^2$Istituto di Astrofisica Spaziale, C.N.R.,
	    Via Fosso del Cavaliere, Roma, Italy}
\date{}

\pubyear{2002}

\begin{document}

\maketitle

\label{firstpage}

\begin{abstract}
In this letter we present analysis of BeppoSAX data from a long look
at the Seyfert 1 galaxy NGC~7469.
The presence of a soft excess below 0.8 keV is confirmed by our analysis 
and no warm absorber component is required to fit the spectrum.
A complex iron emission line and the Compton reflection hump are clearly 
detected. 
The profile of the line is too broad to associate this feature with 
 distant matter.
In addition, the observed soft excess and the energy of the iron line
E$_{Fe}$=6.8 keV strongly support a scenario in which the hard X-rays are 
reprocessed by a photoionized accretion disc.
This hypothesis was tested fitting the BeppoSAX spectrum with the 
ionized disc reflection model of Ross \& Fabian.
A second narrow line component, in addition to that produced in the disc, 
is also required to fit the observed iron line profile.
A high energy cut-off around 150 keV is clearly detected in the spectrum.
\end{abstract}


\section[Introduction]
{Introduction}

A narrow iron line component is clearly detected in 
recent Chandra (Kaspi \etal 2001, Yaqoob \etal 2001) and XMM observations
(Reeves \etal 2001, Pounds \etal 2001) of some Seyfert 1 galaxies. 
These observations suggest that part of the Fe line is produced  
very far from the accretion disc. 
A strong broad iron line component is found in the 
X-ray spectra of many Seyfert galaxies (Fabian \etal 2000) 
by ASCA (Nandra \etal 1997, Tanaka \etal 1995, Yaqoob \etal 2002), XMM 
(Wilms \etal 2001) and BeppoSAX (Guainazzi \etal 1999, De Rosa et al. 2002).

The near (z$\sim$0.017) Seyfert 1 \74 shows a very complex X-ray spectrum.
EXOSAT observed a soft excess in this source (Barr 1986)  
which was confirmed by Einstein (Turner \etal 1991). 
Two ROSAT observations in 1991 (Turner \etal 1993) and 1992 
(Brandt \etal 1993) did not resolve the origin of the soft 
spectral component (e.g. a soft excess, a warm absorber or a combination 
of the two).
No evidence of a warm absorber was found in the 1993 ASCA 
spectrum (Guainazzi \etal 1994).
An iron emission line and a flattening of the spectrum above 10 keV were 
detected by Ginga (Piro \etal 1990).
When the spectrum above 10 keV was reproduced with a cold Compton 
reflection component (George \& Fabian 1991; Matt, Perola \& Piro 1991), 
the intrinsic photon index was $\Gamma$= 1.99 $\pm^{0.06}_{0.05}$ 
(Nandra 1991), while a partial 
covering model yielded $\Gamma$= 1.92 $\pm$ 0.03 (Piro \etal 1990).
The line profile was found to be narrow ($\sigma <$ 150 eV) in the ASCA 
spectrum (Guainazzi \etal 1994), suggesting that the
reprocessing material has to be far from the central source (several hundreds 
of Schwarzschild radii).
Nevertheless Nandra \etal (1997) found marginal evidence of a broad 
component of the iron line in analysing the ASCA spectrum.

The broad band of BeppoSAX (Boella \etal 1997) is particularly suited to 
the deconvolution of such complex spectra.
In this letter we present the analysis of a BeppoSAX week-long observation of
\74 in 1999. 
This spectrum was already analysed by Perola \etal (2002) within 
a sample of nine sources observed with BeppoSAX. 
Their analysis showed evidence of soft excess component and a not resolved 
iron line was also detected, indicating a component from the centre of the 
accretion flow.

\section[Observation]
{Observation}

BeppoSAX observed \74 from 1999 November 23 to 1999 November 29. 
The net exposure time in the MECS was 249180 s. 
The observed mean flux was 
F$_{2-10 keV}=3.7\times 10^{-11}$erg cm$^{-2}$ s$^{-1}$ for our best fit model
(see Table \ref{table_line1}).
Spectra were extracted within a circular region centered on the 
source with radii of 4' and 6' for MECS and LECS respectively.
The background was extracted from event files of source-free regions 
(``blank fields'').
The PDS spectrum was filtered with fixed rise time.
The BeppoSAX data were then fitted using the XSPEC 11.1 package.
All quoted uncertainties correspond to 90 per cent confidence intervals for
one interesting parameter ($\Delta\chi^2$=2.71).
Each model we tested was multiplied by a normalization constant
in order to take into account possible miscalibrations between the different 
instruments. We allowed the PDS/MECS normalization to vary between 
0.77 and 0.95, while the LECS/MECS normalization ratio was
running between 0.7 and 1 (Fiore, Guainazzi \& Grandi 1999).

\begin{figure}
\psfig{figure=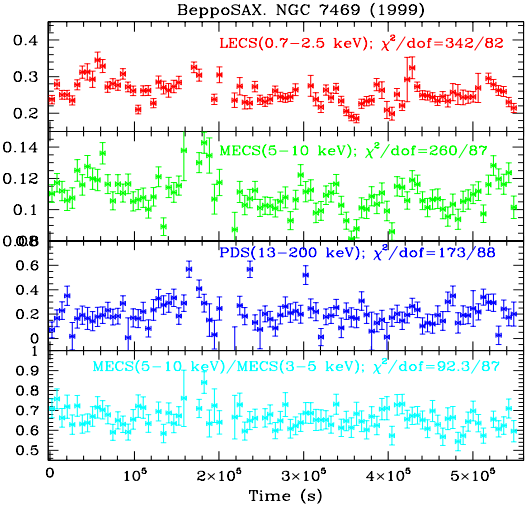,width=0.45\textwidth,height=0.4\textwidth}
\caption{LECS, MECS, PDS lightcurves and MECS hardness ratio 
of BeppoSAX observation of \74. In each panel is indicated
the $\chi^2$/dof for a constant hypothesis.}  
\label{light}
\end{figure}

In Figure \ref{light} we show the LECS (0.7-2.5 keV), 
MECS (5-10 keV) and PDS (13-200 keV) lightcurves, the hardness ratio 
MECS (5-10 keV) / MECS (3-10 keV) is also plotted in the last panel.
In each plot we reported the $\chi^2/$ dof for a constant 
hypothesis.
Even if the source showed flux variability in the soft, medium and hard 
energy range (P$_{\chi^2} \ll$ 10$^{-3}$ for LECS, MECS and PDS lightcurves),
the MECS hardness ratio (which is in the energy range of interest to 
investigate the Fe line behaviour) is in good agreement with a constant value 
(P$_{\chi^2}$= 0.33).
A detailed discussion about the complex spectral variability in 
NGC~7469 (Nandra \& Papadakis 2001 and references therein), is beyond 
the focus of this letter. 

\begin{figure}
\psfig{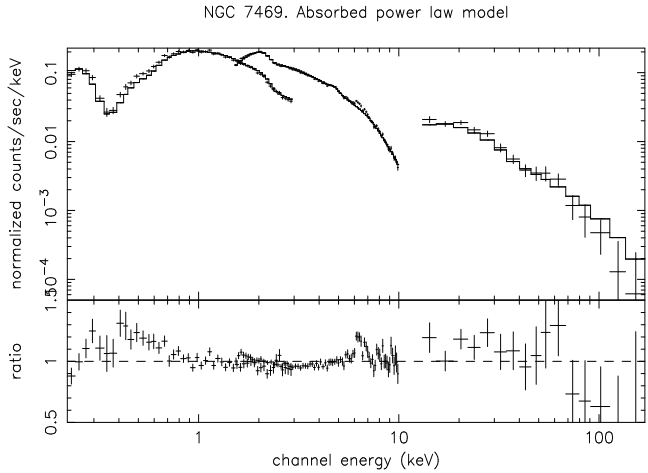}
\caption{LECS, MECS and PDS data (upper panel), and data/model ratio
(lower panel) in the case of an emission continuum fitted with a
power law absorbed by a cold galactic gas.}
\label{spo}
\end{figure}

\begin{figure}
\psfig{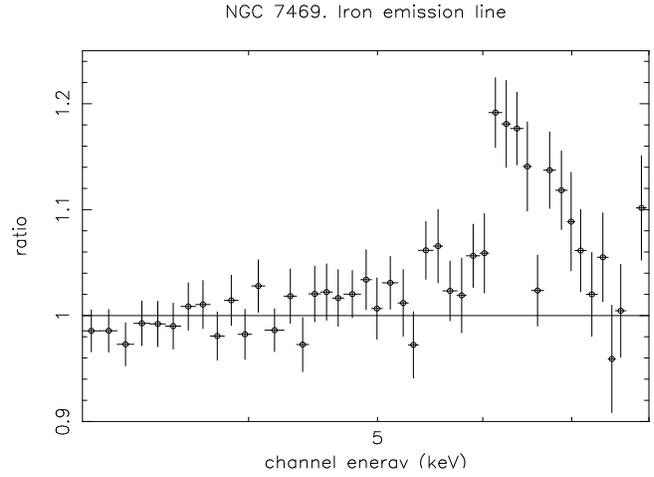}
\caption{Ratio data/model of the iron line when the data are 
fitted with the baseline model in the whole BeppoSAX band except the 
4-7.5 keV energy range.}
\label{line1}
\end{figure}

\begin{figure}
\psfig{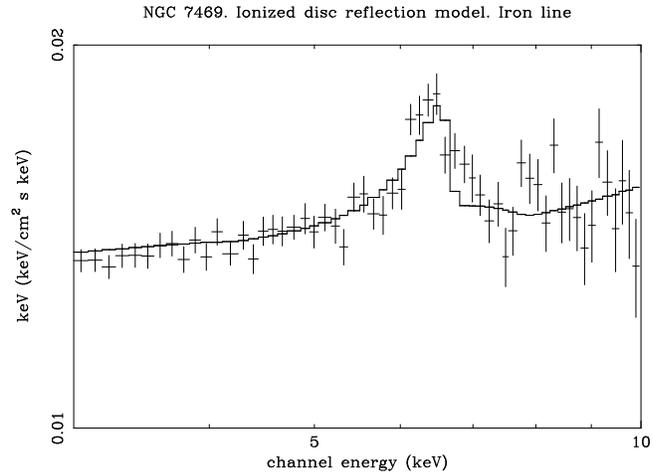}
\caption{The unfolded BeppoSAX/MECS spectra of \74 is plotted with the 
ionized disc reflection model with solar Fe abundance 
(see Table \ref{table_iondisc}). Clear residuals around the Fe line can be 
observed.}
\label{ionline1}
\end{figure}

\begin{figure}
\psfig{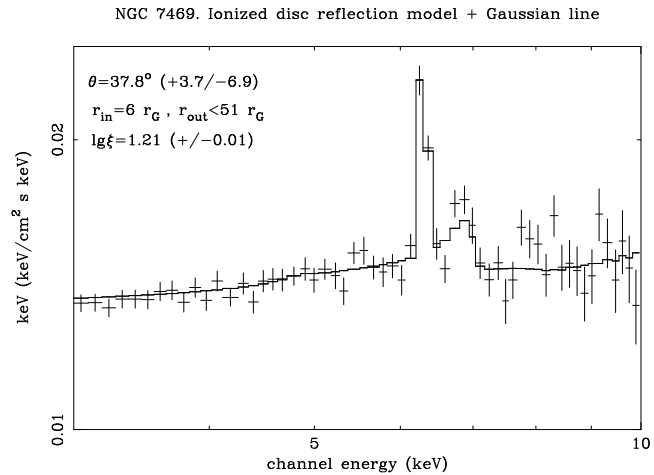}
\caption{A similar plot to Figure \ref{ionline1} but with the spectrum fitted
with an ionized reflection model plus a gaussian component.}
\label{ionline2}
\end{figure}


\begin{table*}
\caption{Cold disc reflection model. $^{(\flat)}$ The inclination angle 
is linked to that of the reflection component and frozen to 30$^\circ$.
$^{(\bullet)}$ Gravitational radius R$_G$=GM/c$^2$.
$^{(\star)}$ In this model for the narrow component E$_L$=6.3 keV 
and $\sigma$=0 keV. All the line energies are quoted in the source rest frame.}
\begin{flushleft}
\begin{tabular}{l c c c c c c c c}
\noalign{\hrule}
\noalign{\medskip}
Model & kT$_{BB}$ & $\Gamma$ & R & E$_{cut-off}$ & E$_L$ & $\sigma$ or R$_{out}$ & EW & $\chi^2/dof$ \\ 
 & (eV) & & & (keV) & (keV) & (keV) or ($^\bullet$R$_G$) & (eV) &  \\
\hline
Single gaussian & 80$\pm^8_4$ & 2.07$\pm^{0.02}_{0.03}$ & 1.49$\pm$0.40 & 245$\pm^{693}_{117}$ & 6.42$\pm^{0.11}_{0.12}$ & $>$0.39 & 175$\pm$80 & 151.3/136 \\ 
 & & & & & & & & \\ 
Diskline$^\flat$ & 84\p9 & 2.06$\pm$0.02 & 1.4$\pm^{0.2}_{0.3}$ & 260$\pm^{630}_{100}$ & 6.60$\pm^{0.25}_{0.16}$ & $>$ 52 & 180\p70 & 155.5/136 \\
 & & & & & & & & \\ 
Diskline$^\flat$ + Narrow gaussian$^\star$ & 84$\pm^7_9$ & 2.06\p0.02 & 1.8$\pm^{0.4}_{0.6}$ & 140$\pm^{385}_{33}$ & 6.82$\pm$ 0.15 & $>$ 15 & 121 $\pm$ 100 & 142.9/134 \\ 
 & & & & & & & 75 $\pm$ 40 & \\ 
\hline
\end{tabular}
\end{flushleft}
\label{table_line1}
\end{table*}

\begin{table*}
\caption{Ionized disc reflection model. The exponent for the disc 
emissivity law r$^\beta$ is frozen to the value $\beta$=-3. 
$^{(\dag)}$ Parameter fixed duringthe fit. $^{(\ddag)}$ 
Parameter pegged at upper limit.}
\begin{flushleft}
\begin{tabular}{l  c c c c c c c c}
\noalign{\hrule}
\noalign{\medskip}
Model & $\Gamma$ & log $\xi$ & R & $\theta$ & $^\dag$r$_{in}$ & r$_{out}$ & E$_L$ & $\chi^2/dof$ \\ 
& & & & ($^{\circ}$) & (R$_G$) & (R$_G$) & (keV)  & \\
\hline
 & & & & & & & \\
Ionized disc & 2.01$\pm$ 0.02 & 1.08$\pm^{0.19}_{0.08}$ & 0.95$\pm^{0.09}_{0.04}$ &20.6$\pm^{6.6}_{8.6}$ & 6 & $^\ddag$10$^7$ & -  & 169/139 \\
 & & & & & & & \\
Ionized disc + gaussian & 2.02$\pm$ 0.02 & 1.21$\pm$0.02 & 0.87$\pm^{0.08}_{0.10}$ & 37.8$\pm^{3.7}_{6.9}$ & 6 & $<$ 51 & 6.30$\pm$ 0.05 & 151.9/137 \\
\hline
\hline
 & & & & ASCA & & & \\
\hline 
& & & & & & & \\
Ionized disc + gaussian & 2.02 $\pm$ 0.03 & $1.08\pm^{0.42}_{0.08}$ 
& 1 (frozen) & $<$40 & 6 & $^\ddag$10$^7$ & 6.36$\pm$0.06& 399.3/378  \\
\hline
\end{tabular}
\end{flushleft}
\label{table_iondisc}
\end{table*}


\section[Results]
{Results}

We fitted the LECS (0.2-3 keV), MECS (1.5-10 keV) and PDS (13-200 keV) 
data simultaneously.
In Figure \ref{spo} we show the data and data/model ratio when the whole 
BeppoSAX spectrum is fitted by an absorbed 
($N_H=4.8\times10^{20}cm^{-2}$ the 
Galactic value from Elvis \etal 1989) power law.
This model is clearly inadequate, with several features emerging from
the spectra: an excess at energy less than 1 keV, an iron line component
with the Compton reflection hump, and a deficit of counts above 70 keV.

The baseline model we employed to fit the data consisted of an absorbed 
cut-off power law, a cold reflection component (PEXRAV in XSPEC 
Magdziarz \& Zdziarski 1995) and a black body to reproduce the excess at 
low energy.
The probability to exceeding F with the addition of the high energy 
cut off in the baseline model is P$_F \sim$ 96.5 per cent.
In Figure \ref{line1} the MECS data/model ratio is shown 
when the baseline model is applied to the whole BeppoSAX band, 
but excluding 4-7.5 keV energy range where a relativistic iron line 
is expected to be present.
The ratio in Figure \ref{line1} shows clearly the presence of a complex 
iron line.
We tried several models to fit the line profile and all the best fit 
parameters are shown in Table \ref{table_line1}.
A single gaussian is clearly inadequate 
($\chi^2/dof$=151.3/136).
The second step was to substitute an unphysical broad gaussian component 
with a physical model for a relativistic profile of the line 
(DISKLINE in XSPEC, Fabian \etal 1989).
The model is still inadequate to reproduce the line residuals 
($\chi^2$/dof=155.5/136).
The addition of a second narrow gaussian component at the energy of the 
neutral iron yields a significant decrease of $\chi^2$ ($\Delta\chi^2$=13 with
the addition of 2 free parameters. P$_F$=99.70 per cent). 
The EW was 121 $\pm$ 100  eV and 75 $\pm$ 40 eV for the broad and narrow 
components respectively. The centroid of the relativistic line 
E$_L$=6.82 $\pm$ 0.15 keV, indicates He-like iron.

Two gaussian narrow lines give a fair $\chi^2$ ($\chi^2$/dof=142/135).
The second line (EW=57$\pm$36 eV) lies to the energy 
of highly ionized iron E=6.94$\pm^{0.21}_{0.12}$ keV, i.e. H-like iron.
This component has to be produced in a gas characterized by 
an ionization parameter $\xi=4\pi L_{ion}/n_e R^2 >$ 2000 erg cm s$^{-1}$ 
(Matt \etal 1996).
Assuming a temperature T=10$^6$ K, the total recombination rate to all states 
for FeXXVI is $\alpha_{tot}=2.4\times 10^{-11}$ cm$^3$ s$^{-1}$ 
(Seaton 1959), and the line emissivity is  
$J_{line}=\frac{n_e^2}{4\pi} A_{Fe}\,y\,h\nu\,\alpha_{tot}$, with $A_{Fe}$ 
and $y$ the iron aboundance and the iron effective fluorescent 
yield respectively. 
In the case of a simple spherical geometry if we assume a ionizing 
luminosity $L_{ion}\sim$ 10$^{43}$ erg s$^{-1}$, and with the observed 
H-like iron line luminosity $L_{line}\sim$ 10$^{41}$ erg s$^{-1}$, 
we can estimate a lower limit to the electron number density 
$n_e > (\frac{3 L_{line}} {A_{Fe} y E_{line}\alpha_{tot}})^2 
\times (\frac{L_{ion}}{\xi})^{-3}=2.4 \times 10^{10}$ cm$^{-3}$ (assuming 
$A_{Fe}=3.3\times 10^{-5}$ of Morrison \& McCammon 1983 and $y\sim 0.63$ 
by Matt \etal 1993b for a temperature of 10$^6$ K)    
and then an upper limit for the distance of the gas from the ionizing 
central source, R $< 7\times 10^{14}$ cm.
This means that such gas must lie within few hundred Schwarzschild radii 
(R$_s$=2GM/c$^2$) of the central black hole 
(assuming a mass for it of $\sim$ 10$^7$ M$_\odot$, Wandel \etal 1999).

The fit with two gaussian emission lines (one narrow) is still 
perfectly acceptable ($\chi^2$/dof=133/134), nevertheless the extremely 
large value of the intrinsic width of the broad component 
($\sigma$=1.8$\pm^{1.0}_{0.7}$ keV) indicates a poor description 
of the continuum at the energy of the line.
Thus we are lead to believe that a diskline + gaussian component is 
a more suitable and physically consistent model to reproduce 
the Fe emission line observed in \74.

The presence of an ionized iron line and of a soft emission component, 
together with an unusual value of the relative reflection, significant greater 
than 1 (R=1.8$\pm^{0.4}_{0.6}$, see Table \ref{table_line1}), strongly 
suggests the hypothesis that all these features could be the 
result of hard X-ray reprocessing in a hot ionized accretion disc 
(Matt \etal 1993a, 1996, Ross, Fabian \& Young 1999, Nayakshin \etal 2000,
Ballantyne \etal 2001). 
We tested an ionized reflection model which takes into account the reflected 
spectrum of the disc in the whole BeppoSAX energy range.
The model we employed was that of Ross \& Fabian (1993). 
The most important quantity in determining the shape of the reflected continuum
is the ionization parameter $\xi=4\pi F_X/n_H$, where F$_X$ is the X-ray flux 
(between 0.1-100 keV) illuminating a slab of gas with solar abundances and 
constant hydrogen number density n$_H$=10$^{15}$ cm$^{-3}$.
The incident flux is assumed be a power law with spectral photon index 
$\Gamma$ and a high energy cut-off at 100 keV. The computed reflected 
spectrum is multiplied for a factor R (``reflected fraction'') and added 
to the primary continuum. 
This model includes the Fe K$\alpha$ emission line and the spectral 
features at low energy (emission lines and recombination continuum).
Larger values of $\xi$ indicates a more ionized 
disc which will affect the strength and width of the iron line 
(Matt \etal 1993a, 1996), and the absorption edges. The soft emission from 
the accretion disc is not take into account to determine the reprocessing 
features. 
During the fit we applied to the spectrum a 
relativistic blurring appropriate for a Schwarzschild geometry assuming a 
disc emissivity law (Fabian \etal 1989). 

Fitting the spectrum with an ionized disc model (see Table \ref{table_iondisc})
with solar Fe abundance does not give a satisfactory result 
($\chi^2/dof$=169/139). 
Clear residuals can be observed at the energy of the iron line 
(see Figure \ref{ionline1}). 
To increase the Fe emissivity to reproduce the observed strength of the 
line we tried to fit the data with an ionized reflection model with 2 and 
5 times solar Fe abundances. Nevertheless these models are inadequate to 
fit the soft part of the spectrum, and the result of the fit is poor
($\chi^2/dof_{(2XFe)}$=187/139, $\chi^2/dof_{(5XFe)}$=205/139).
A better model to reproduce the iron line residuals is to add a narrow line 
component. The ionized disc model with an additional gaussian component finally
give us a good fit ($\Delta\chi^2$=17 for the sum of two interesting 
parameters. P$_F$=99.97 per cent) and no large residuals can be found at 
the energy of the iron line (see Figure \ref{ionline2}).
The best fit parameters for the several ionized reflection models we employed
are shown in Table \ref{table_iondisc}. In Figure \ref{reflection} we plot
the confidence contours (68, 95 and 99 per cent) Reflection fraction 
and photon index for the ionized disc plus gaussian line model.

No additional (cold or ionized) iron edge is required to fit the spectrum.
If we add an absorption edge with the energy fixed to the value for 
cold iron E$_{edge}$=7.1 keV, the fit worsens
and we find a very low upper limit for the optical depth $\tau<$ 0.01.

The soft excess at low energy can not be accounted for by a warm absorber 
component. If we substitute the black body in the last model 
in Table \ref{table_line1}, with two absorption edges at the energy of 
the OVII (0.74 keV) and OVIII (0.87 keV), the fit is worse 
($\chi^2/dof$=150/134).
The same result is obtained using a more detailed description of the 
warm absorber with ABSORI model in XSPEC. 
A warm absorber component is not required 
to fit the low energy data of NGC~7469 (P$_F$=34 per cent).

Piro \etal (1990) showed that the GINGA spectrum of \74 observed in 1988 
could be reproduced by a ``partial covering'' model (Holt \etal 1980). 
In this scenario an absorbing medium 
is supposed to cover only a fraction $f_{cov}$ of the central source with 
a column density N$_{Hcov}$. The direct X-ray would be blocked at low energy, 
but would penetrate through the absorbing gas above 5 keV imprinting the 
curvature in the observed spectrum. 
Nevertheless Leighly \etal (1996) showed that this model was not able to 
explain the observed flux and spectral variability in \74.

When we model the BeppoSAX spectrum of \74 with partial covering 
the fit is fair ($\chi^2/dof$=142.7/136), with 
N$_{Hcov}=(5 \pm 1)\times 10^{23}$ cm$^{-2}$ and $f_{cov}$= (0.32 $\pm$ 0.04),
and the EW of the iron line EW=69 $\pm$ 20 eV.
Nevertheless the expected value of the equivalent width for an optically thin 
medium illuminated by an isotropic continuum, with a Fe solar abundance and 
the  N$_{Hcov}$ and $f_{cov}$ obtained by BeppoSAX observation, is 
in the range 100-200 eV, i.e. larger than that observed in the 
spectrum.
This evidence together with the complex profile of the Fe line and the 
soft X-ray excess, make the reflection from a (possibly ionized) accretion 
disc the most appealing scenario.

\section[Discussions and conclusions]
{Discussions and conclusions}

The BeppoSAX long look at \74 shows a very complex X-ray spectrum.
Both a soft excess component and reprocessing features are detected.
The continuum emission can be reproduced either with a cold reflection model 
and a black body component (with temperature kT $\sim$ 84 eV fully consistent 
with that measured by ASCA, Guainazzi \etal 1994) or with an ionized disc 
reflection model that takes into account the disc emissivity below 0.8 keV.
Whichever model we employ to fit the spectrum, the observed iron 
line requires a complex profile which is hard 
to associate with a distant reflector.
In addition the energy of the relativistic line E=6.8 keV corresponds 
to high ionized iron that can not be reconciled with the cold disc scenario.

\begin{figure}
\psfig{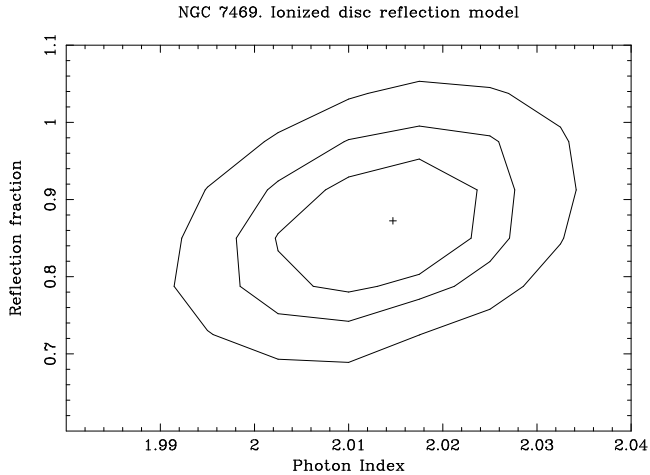}
\caption{68, 95 and 99 per cent confidence contours Reflection fraction 
$vs$ photon index for the ionized disc plus gaussian line model.The ``+'' 
denotes the best fit values of R and $\Gamma$.}
\label{reflection}
\end{figure}

An ionized iron line component was already observed by XMM in Mkn~509 
(Pounds \etal 2001), Mkn~205 (Reeves \etal 2001) and 
NGC~5506 (Matt \etal 2001).
In this last source a solution in term of a blend of He- and H- like narrow 
iron line is preferable to that of a relativistic ionized disc, but in 
Mkn~509 and Mkn~205 this feature was explained in terms of an ionized 
accretion disc.
In the case of \74 the predicted iron line component from the ionized disc 
model does not fit the whole observed profile.
A narrow line component has to be included in the model.
The origin of this second component is still a matter of debate.
It has been detected in many Seyfert 1s observed so far by 
Chandra (Kaspi \etal 2001, Yaqoob \etal 2001) and XMM (Pounds \etal 2001, 
Reeves \etal 2001). It is associated with a region distant from the 
central source and probably coincident either with a Compton thick gas, 
e.g. the ``torus'' obscuring the Seyfert 2s (Antonucci 1993, 
Ghisellini \etal 1994) or with the Compton thin BLR. 
At least in the cases of NGC~4051 (Guainazzi \etal 1998) 
and NGC~5506 (Matt \etal 2001) the narrow core of the iron line was
associated to a Compton thick gas.
Nevertheless the bulk of the Compton reflection in \74 can be completely 
associated with the ionized disc (see Figure \ref{reflection}) iron component.
In fact with the combination of a cold and an ionized Compton reflection 
we obtained a very poor fit ($\chi^2/dof$=156.4/135).

Even if a partial covering model can reproduce the observed 
BeppoSAX spectrum, the value of the equivalent width of the iron 
line predicted in this model is significantly larger than that measured.
This could require that the covering medium has a small solid angle at 
the source.

A warm absorber component is not required to fit the BeppoSAX low energy 
spectrum, this allowed us to detect the redward extent of the iron line, 
and thus constrain the inner radius of the disc.

A double iron line component also can reproduce the observed profile. 
We cannot rule out the possibility that the second H-like iron component 
is created in a warm (T $\sim 10^6$ K) gas within a few 100s R$_s$ of 
the central source.

Finally to check our results we fitted the the ASCA
spectrum above 1 keV 
(SIS0 and GIS2, sequence number 71028030, from the Tartarus database) 
with an ionized disc reflection model. 
The fit was good ($\chi^2/dof$=399.3/378) and the ionization parameter 
was consistent with that observed in the BeppoSAX spectrum, 
log ${\xi}_{ASCA}=1.08\pm^{0.42}_{0.08}$.
A narrow line component also is really marginally required 
(P$_F$=92 per cent) to fit the Fe line. 
The flux of this component was the same in the two observations, while the 
flux in 2-10 keV was higher by 30 per cent during the SAX long look.
Nevertheless the uncertainty on the line intensity rends  
any result about variability inconclusive.

\section*{Acknowledgments}
We thank David Ballantyne for computing the grid of ionized
disc models.
A.D.R acknowledges  the European Association for Research 
in Astronomy (EARA) Marie Curie for financial support.
The BeppoSAX satellite is a joint Italian-Dutch program.

\bsp

\label{lastpage}


\begin{thebibliography}{}

\bibitem[Antonucci 1993]{antonucci93}
Antonucci R., 1993, ARA\&A, 31, 473

\bibitem[Ballantyne \etal 2001]{ballantyne01}
Ballantyne D., Ross R.R., Fabian A.C., 2001, MNRAS, 327, 10

\bibitem[Barr 1986]{barr86}
Barr P., 1986, MNRAS, 223, 29p

\bibitem[Boella \etal 1997]{boella97}
Boella G., Butler R.C., Perola G.C, Piro L., Scarsi L., Bleeker J., 
1997, A\&A 122, 299

\bibitem[Brandt \etal 993]{brandt93}
Brandt W.N., Fabian A.C. Nandra K., Tsuruta S., 1993, MNRAS, 265, 996

\bibitem[De Rosa \etal 2002]{derosa2002}
De Rosa A., Piro L., Fiore F., Grandi P., Maraschi L., Matt G., Nicastro F., Petrucci P.O., 2002, A\&A, 38, 838

\bibitem[Elvis \etal 1989]{elvis89}
Elvis M., Lockman F.J., Wilkes B. 1989, AJ, 97, 777

\bibitem[Fabian \etal 1989]{fabian89}
Fabian A.C., Rees M.J., Stella L., White N.E., 1989, MNRAS, 238, 729

\bibitem[Fabian \etal 2000]{fabian2000}
Fabian, A.C., Iwasawa K., Reynolds C.S., Young, A.J., 2000, PASP, 112, 1145

\bibitem[Fiore, Guainazzi \& Grandi 1999]{cook} 
Fiore F., Guainazzi G., Grandi P. 1999, {\it Cookbook for BeppoSAX NFI 
Spectral analysis}. SDC report. 
(Available from http://asdc.asi.it/bepposax/software/index.html)

\bibitem[George \& Fabian 1991]{gf91}
George I.M. \& fabian A.C., 1991, MNRAS, 249, 352

\bibitem[Ghisellini \etal 1994]{ghisellini94}
Ghisellini F., Haardt F., Matt G., 1994, MNRAS, 267, 743

\bibitem[Guainazzi \etal 1994]{guaina94}
Guainazzi M, Matsuoka M., Piro L., Mihara T., Yamauchi M., 1994, ApJ, 436, L35

\bibitem[Guainazzi \etal 1998]{guainazzi98} 
Guainazzi M., Nicastro F., Fiore F., Matt G., McHardy I., Orr A., Barr P.,
Fruscione A., Papadakis I., Parmar A.N., Uttley P., Perola G.C., Piro L., 
1998, MNRAS, 301, L1

\bibitem[Guainazzi \etal 1999]{guainazzi99}
Guainazzi M., Matt G., Molendi S., Orr A., Fiore, F., Grandi P., Matteuzzi, A.,
Mineo, T., Perola G.C., Parmar A.N., Piro L., 1999 A\&A, 341, L27

\bibitem[Holt et al. 1980]{holt80}
Holt S.S., Mushotzky R.F., Boldt E.A., Serlemitsos P.J., Becker R.H., 
Szymkowiak A.E., White N.E., 1980, ApJ, 241, L13

\bibitem[Kaspi et al. 2001]{kaspi2}
Kaspi S., Brandt W.N., Netzer H., George I., Chartas G., Behar E., 
Sambruna R., Garmire G. and Nousek J. 2001, ApJ, 554, 216

\bibitem[Leighly et al. 1996]{leighly96}
Leighly K., Kunieda H., Awaki H., Tsuruta S., 1996 ApJ, 463, 158


\bibitem[Magdziarz \& Zdziarski 1995]{MZ95}
Magdziarz P. \& Zdziarski A.A. 1995, MNRAS, 273, 837

\bibitem[Matt, Perola \& Piro 1991]{MPP91}
Matt G., Perola G.C., Piro L. 1991, A\&A, 247, 25

\bibitem[Matt \etal 1993a]{MFR93}
Matt G., Fabian A.C., Ross R.R., 1993a, MNRAS, 262, 179

\bibitem[Matt \etal 1993b]{MBF93}
Matt G., Brandt W.N. and Fabian A.C., 1993b, MNRAS, 280, 823

\bibitem[Matt \etal 1996]{MFR96}
Matt G., Fabian A.C., Ross R.R., 1996, MNRAS, 278, 1111

\bibitem[Matt \etal 2001]{matt01}
Matt G.,  Guainazzi M., Perola G. C.,Fiore F., Nicastro F.,Cappi M.,
Piro L., 2001b, A\&A, 377, L31

\bibitem[Morrison \& McCammon 1983]{}
Morrison R. \& McCammon D.,  1983, ApJ, 270, 119
 
\bibitem[Nandra 1991]{nandra91}
Nandra K., 1991, PhD thesis, Leicester University

\bibitem[Nandra \etal 1997]{nandra97}
Nandra, K., George I.M., Mushotzky R.F., Turner T.J., Yaqoob, T., 1997, ApJ, 
477, 602

\bibitem[Nandra \& Papadakis 2001]{nandra01}
Nandra K. \& Papadakis I.E., 2001, ApJ, 554, 710

\bibitem[Nayakshin \etal 2000]{nayakshin00}
Nayakshin S., Kazanas D., Kallmann T., 2000, ApJ, 537, 833 

\bibitem[Perola \etal 2001]{P2002}
Perola G.C., Matt G., Cappi M., Fiore F., Guainazzi M., Maraschi L., 
Petrucci P.O., L. Piro, 2002, A\&A, accepted, astro-ph/0205045

\bibitem[Piro \etal 1990]{piro90}
Piro L, Yamauchi M. \& Matsuoka M., 1990, ApJ, 360, L35

\bibitem[Pounds \etal 2001]{pounds2001}
Pounds K., Reeves J., O'Brien P., Page K., Turner M., Nayakshin S., 2001,
ApJ, 559, 181

\bibitem[Reeves \etal 2001]{reeves2001}
Reeves J.N., Turner M., Pounds K., O'Brien P., Boller Th., Ferrando P., 
Kendziorra E., Vercellone S., 2001, ApJ, 365L, 134

\bibitem[Ross, Fabian \& Young 1999]{RFF99}
Ross R.R., Fabian A.C., Young A.J., 1999, MNRAS, 306, 461

\bibitem[Rees \& Fabian 1993]{RF93}
Ross R.R. \& Fabian A.C., 1993 MNRAS, 261, 74

\bibitem[Seaton 1959]{seaton59}
Seaton M.J., 1959, MNRAS, 119, 81.

\bibitem[Tanaka \etal 1995]{tanaka95}
Tanaka Y, Nandra K., Fabian A.C., Inoue H., Otani C., Dotani T.,
Hayashida K., Iwasawa K., Kii T., Kunieda H., Makino F., 
Matsuoka M., 1995, Nature, 375, 659

\bibitem[Turner \etal 1991]{turner91}
Turner T.J., Weaver K.A., Mushotzky R.F., Holt S.S., Madejeski G.M., 
1991, ApJ, 381, 85

\bibitem[Turner \etal 1993]{turner93}
Turner T.J., George I.M., Mushotzky R.F., 1993, ApJ, 412, 72

\bibitem[Yaqoob \etal 2001]{yaq2001}
Yaqoob, T., George I.M., Nandra K., Turner T.J., Serlemitsos P.J., 
Mushotzky R.F., 2001, ApJ, 546, 759

\bibitem[Yaqoob \etal 2002]{yaq2002}
Yaqoob T., Padmanabhan U, Dotani T, Nandra K., ApJ, accepted. astro-ph/0112318

\bibitem[Wandel \etal 1999]{wandel99}
Wandel A., Peterson B.M. and Malkan M.A., 1999, ApJ, 526, 579.

\bibitem[Wilms \etal 2001]{wilms01}
Wilms J., reynolds C.S., begelman M.C., Reeves J., MOlendi S., Staubert R., 
Kendziorra E., 2001, MNRAS, 328, L27

\end{thebibliography}
\end{document}